\documentclass[pra,amsmath,amssymb,showpacs,superscriptaddress,twocolumn]{revtex4}
\usepackage{amsmath,mathrsfs,amsbsy,color,graphicx,bm,amsthm,amsfonts}
\usepackage{units}
\usepackage{bbm}
\usepackage{times}
\usepackage{dcolumn}
\usepackage{mathrsfs}
\usepackage{amsmath,amssymb,epsfig,float}

\newcommand{\norm}[1]{\lVert #1 \rVert}
\begin{document}

\title{Irreversible degradation of quantum coherence under  relativistic motion}

\author{Jieci Wang}
\email{jcwang@hunnu.edu.cn}
\affiliation{Department of Physics, Collaborative Innovation Center for Quantum Effects, and Key Laboratory of Low
Dimensional Quantum Structuresand Quantum \\
Control of Ministry of Education,
 Hunan Normal University, Changsha, Hunan 410081, China}
\affiliation{Beijing National Laboratory for Condensed Matter Physics, Institute of Physics,Chinese Academy of Sciences, Beijing 100190, China}

\author{Zehua Tian}

\affiliation{Institute of Theoretical Physics, University of Warsaw, Pasteura 5, 02-093 Warsaw, Poland}

\author{Jiliang Jing}
\email{jljing@hunnu.edu.cn}
\affiliation{Department of Physics, Collaborative Innovation Center for Quantum Effects, and Key Laboratory of Low
Dimensional Quantum Structuresand Quantum \\
Control of Ministry of Education,
 Hunan Normal University, Changsha, Hunan 410081, China}

\author{Heng Fan}
\affiliation{Beijing National Laboratory for Condensed Matter Physics, Institute of Physics,Chinese Academy of Sciences, Beijing 100190, China}

\begin{abstract}

\vspace*{0.2cm} We study the dynamics of quantum coherence under Unruh thermal noise and seek under which condition the coherence can be frozen in a relativistic setting. We find that  the frozen condition is either (i) the initial state is prepared as a incoherence state, or (ii) the detectors have no interaction with the external field.  That is to say, the decoherence of detectors' quantum state is  irreversible under  the influence of thermal noise induced by Unruh radiation. It is shown that quantum coherence  approaches zero only in the limit of an infinite acceleration, while quantum entanglement could reduce to zero for a finite acceleration. It is also demonstrated that the robustness of quantum coherence is better than entanglement  under the influence of the atom-field interaction for an extremely large acceleration. Therefore, quantum coherence is more robust than entanglement in an accelerating system and the coherence type quantum resources are more accessible for relativistic quantum information processing tasks.
\end{abstract}

\pacs{04.62.+v, 03.70.+k, 06.20.-f, 05.30.Jp }

\maketitle
\section{Introduction.}

It is well known that quantum mechanics and  relativity theory are two  fundamentals of modern physics.  Since the early 20th century, much effort have been put forward to bridge the gap between
them. Reconciling quantum mechanics with general relativity gave birth to quantum field theory (QFT) and several predictions have been made based on the QFT. An important prediction in QFT is the   Unruh effect \cite{unruh1976, unruhreview},
which  tells us that
a uniformly accelerated observer in Minkowski spacetime observes the Minkowski vacuum of quantum fields as a
thermal bath. The Unruh effect reveals that the concept of particle is observer-dependent and the temperature
of the thermal bath is proportional to the proper acceleration
of the observer. Recently, a number of authors considered the influence of the Unruh effect on quantum systems in an accelerated setting  \cite{teleport,Schuller-Mann,RQI1,RQI3,RQI4,
RQI5,Landulfo,Landulfo1,tianwang,wangtian,hujhep}.  It has been found that the quantity of quantum resources, such as entanglement, discord-type quantum correlation and nonlocality, suffer from a degradation from the viewpoint of noninertial observers due to the effect of the Unruh thermal bath.

On the other hand, quantum coherence, arising from the quantum
state superposition principle, is one of the key fundamental aspects of quantum physics ~\cite{Leggett}.  Analogous to entanglement and discord-type quantum correlation,  coherence  is regarded as a physical resource in quantum optics experiments~\cite{Glauber,Scully,Albrecht,Walls},  as well as various quantum information processing  tasks~\cite{Nielsen, Baumgratz}.
Despite the fundamental importance of quantum coherence, it received
increasing attention until Baumgratz \emph{et. al.} introduced a rigorous framework for the measurement of
coherence and proposed several quantifiers, such as the relative entropy of coherence and the \textit{$l_1$} norm ~\cite{Baumgratz}. Recently, Streltsov \emph{et. al.} found that any degree of coherence with respect to some
reference basis can be
converted to entanglement via incoherent operations \cite{CoherenceWithEnt}.
 Shao \emph{et. al.} proved that the fidelity does not  satisfy the monotonicity requirement as a measure of coherence
 while the  trace norm of coherence \cite{FidelityTrace} is a  promising candidate for coherence monotone.
Most recently, it has been found that  the coherence of an
open system  can be  frozen under some dynamical conditions~\cite{thomas, xiaobao}.

In this paper we study quantum coherence between a pair of  Unruh-Dewitt detectors \cite{UW84} when one of them is accelerated. The detectors are modeled by  two-level semiclassical atoms with  fixed energy gap and are designed to interact locally with the neighbor scalar fields. We assume that Alice's detector is always switched off and  remains stationary while Bob's detector moves with a constant acceleration and interacts with the massless scalar field. The quantum resource aspects,  such as entanglement \cite{Landulfo}, discord-type quantum correlations \cite{Landulfo1}, and quantum nonlocality \cite{tianwang} of two entangled detectors have been discussed recently. The advantage of entangled detectors for quantum metrology \cite{wangtian} under the influence of the Unruh effect has also been studied. The detector is \emph{classical} in the sense that it possesses a classical world line  and is\emph{quantum} because its internal degree of freedom is treated quantum mechanically \cite{UW84,Landulfo}.
 We find that the quantum coherence can not be frozen for any acceleration due to the effect of Unruh thermal noise. In addition, quantum coherence is found to be  more robust than entanglement under Unruh decoherence and therefore the coherence type quantum resources are more accessible than entanglement for relativistic quantum information processing tasks. 

The outline of the paper is as follows. In Sec. II we discuss the quantum information description of the entangled Unruh-Dewitt detectors and the evolution of a prepared state in the case of only one
detector with relativistic motion. In Sec. III, we recall the common conditions that any coherence quantifier should satisfy and introduce a few coherence measurements.  In Sec. IV, we study the dynamics of quantum coherence under Unruh thermal noise and seek the conditions
 for the frozen of quantum coherence.
The conclusion are given in the last section.

\section{Evolution of the detectors' state  under  relativistic Motion}

We consider two observers, Alice and Bob, where each of them possesses a Unruh-Dewitt detector \cite{UW84} modeling through a two-level non-interacting atom \cite{Landulfo,Landulfo1,tianwang}. The atom carried by Alice keeps static and  Rob's detector moves with uniform acceleration $a$ for a
time duration $\Delta$. Then we let
Alice's detector always be switched off and the one holed by Rob is switched on when it moves with  constant acceleration. The world line of Rob's detector is
\[
t(\tau)=a^{-1}\sinh a\tau,\;x(\tau)=a^{-1}\cosh a\tau,
\]
$y(\tau)=z(\tau)=0$, where $a$ is Rob's proper acceleration and $\tau$ is the proper time of the detector. Throughout this paper, we set $c=\hbar=\kappa_{B}=1$. We assume that the initial
state of the detector-field system has the form
\begin{equation}
|\Psi_{t_0}^{AR\phi}\rangle=|\Psi_{AR}\rangle\otimes|0_{M}\rangle
,\label{IS}%
\end{equation}
where $|\Psi_{AR}\rangle=\sin\theta |0_{A}\rangle|1_{R}\rangle+\cos\theta|1_{A}\rangle
|0_{R}\rangle$ denotes the initial state shared by Alice's (A) and Bob's (R) detectors, and $|0_{M}\rangle$ represents the external scalar field is in  Minkowski vacuum.

The total Hamiltonian of the  system is given by
\begin{equation}
H_{A\, R\, \phi} = H_A + H_R + H_{KG} +  H^{R\phi}_{\rm int},\label{totalh}
\end{equation}
where $H_{A}=\Omega A^{\dagger}A$ and $H_{R}=\Omega R^{\dagger}R$ are the detectors'  Hamiltonian and $\Omega$ is the energy gap of the detectors.
 The interaction Hamiltonian $H^{R\phi}_{\rm int}(t)$ between the accelerated detector and the external scalar field is given by
\begin{equation}
H^{R\phi}_{\rm int}(t)=
\epsilon(t) \int_{\Sigma_t} d^3 {\bf x} \sqrt{-g} \phi(x) [\chi({\bf x})R +
                           \overline{\chi}({\bf x})R^{\dagger}],
\label{int}
\end{equation}
where $g\equiv {\rm det} (g_{ab})$, and $g_{ab}$ is the Minkowski metric. Here  $\chi(\mathbf{x})=(\kappa\sqrt{2\pi})^{-3}\exp(-\mathbf{x}^{2}/2\kappa^{2})$ is a coupling function which vanishes outside a small volume
around the detector. Such a Gaussian coupling function describes a point-like detector which only interacts with the neighbor scalar fields \cite{KY03} in the Minkowski vacuum.

In the weak coupling case, we can calculate the final state $|\Psi^{R \phi}_{t = t_0+\Delta} \rangle$
of the atom-field system at time $t=t_0+\Delta$  in the first order of perturbation over the coupling constant $\epsilon$ \cite{UW84}. Under the dynamic evolution described by the Hamiltonian given by Eq. (\ref{totalh}), the final state $|\Psi^{R \phi}_{t} \rangle$
  is found to be
 \cite{Landulfo,Landulfo1,tianwang,wangtian,wald94}
\begin{equation}
|\Psi^{R \phi}_{t} \rangle
= [I - i(\phi(f)R + \phi(f)^{\dagger} R^{\dagger}) ] |\Psi^{R \phi}_{t_0} \rangle,
\label{primeira_ordem}
\end{equation}
where
\begin{eqnarray}
\nonumber\phi(f) &\equiv& \int d^4 x \sqrt{-g}\chi(x)f\\&=&i [a_{RI}(\overline{u E\overline{f}})-a_{RI}^{\dagger}(u Ef)],
\label{phi(f)}
\end{eqnarray}
is a field operator which describes the distribution of the external scalar field \cite{Landulfo,wald94}.
In Eq. (\ref{phi(f)}), $f \equiv \epsilon(t) e^{-i\Omega t}\chi ({\bf x})$ is a compact support complex function defined in the Minkowski spacetime, and
$a_{RI}(\overline{u})$ and $a_{RI}^{\dagger}(u)$ are the annihilation and creation
operators of $u$ modes \cite{Landulfo,Landulfo1,tianwang,wangtian,wald94}, respectively.
In addition, $u$ is an operator that takes the positive frequency part of the solutions of the
Klein-Gordon equation in  Rinlder metric \cite{Landulfo,wald94}, and $E$ is the
difference between the advanced and retarded Green functions.

By inserting the  initial state Eq. (\ref{IS}) into Eq. (\ref{primeira_ordem}),
we obtain the final state of the total system in terms of the Rindler operators $a_{R I}^{\dagger}$ and $a_{R I}$, which is given by
\begin{eqnarray}
| \Psi^{AR \phi}_{t}\rangle
& = &
|\Psi^{AR \phi}_{t_0} \rangle
 + \sin \theta |0_A\rangle  |0_R\rangle
 \otimes(a_{R I}^{\dagger}(\lambda)|0_M\rangle)
 \nonumber \\
& + & \cos \theta |1_A\rangle |1_R\rangle\otimes(a_{R I}(\overline{\lambda})|0_M\rangle),
\label{evolutionAUX}
\end{eqnarray}
where  $\lambda = -uEf$. Here the Rindler
operators $a_{R I}^{\dagger}(\lambda)$ and $a_{R I}(\overline{\lambda)}$  are defined in Rindler region $I$, while  the $|0_M\rangle$ is
vacuum state in the Minkowski spacetime.  The Bogoliubov transformations
between the Rindler operators and the operators annihilating the Minkowski $|0_M\rangle$ vacuum state can be written as \cite{Landulfo, wangtian}~
\begin{eqnarray}
a_{R I}(\overline{\lambda})&=&
\frac{a_M(\overline{F_{1 \Omega}})+
e^{-\pi \Omega/a} a_M ^{\dagger} (F_{2 \Omega})}{(1- e^{-2\pi\Omega/a})^{{1}/{2}}},
\label{aniq} \\
a^{\dagger}_{R I}(\lambda)&=&
\frac{a^{\dagger}_M (F_{1 \Omega}) +
e^{-\pi \Omega/a}a_M(\overline{F_{2 \Omega}})}{(1- e^{-2\pi\Omega/a})^{{1}/{2}}}
\label{cria},
\end{eqnarray}
where
$F_{1 \Omega}=
\frac{\lambda+ e^{-\pi\Omega/a}\lambda\circ w}{(1- e^{-2\pi\Omega/a})^{{1}/{2}}}$, and
$F_{2 \Omega}=
\frac{\overline{\lambda\circ w}+ e^{-\pi\Omega/a}\overline{\lambda}}{(1- e^{-2\pi\Omega/a})^{{1}/{2}}}$. Here
$w(t, x)=(-t, -x)$
is a wedge reflection isometry that makes a reflection from $\lambda$ in the Rindler region $I$  to $\lambda\circ w$ in the Rindler region $II$
\cite{Landulfo, wangtian,wald94}.

We are interested in the dynamics of the detectors' state after interacting with the field.  By tracing out the degrees of freedom of the external field $\phi(f)$, we obtain the density matrix that describes the detector's state
\begin{eqnarray}
\rho_{t}^{AR} = \left(
                  \begin{array}{cccc}
                    \gamma & 0 & 0 & 0 \\
                    0 & 2\alpha \sin^2 \theta  & \alpha\sin 2\theta  & 0 \\
                    0 & \alpha\sin 2\theta  & 2\alpha\cos^2 \theta  & 0 \\
                    0 & 0 & 0 & \beta \\
                  \end{array}
                \right)
,
\label{rhof2}
\end{eqnarray}
where $\Psi_{AR}$ is the detectors's initial state in Eq. (\ref{IS}) and the parameters
$\alpha$, $\beta$ and $\gamma$ are given by
\begin{align*}
\alpha  &  =\frac{1-q}{2(1-q)+2\nu^{2}(\sin^2 \theta+q\cos^2\theta)},\\
\beta  &  =\frac{\nu^{2}q\cos^2\theta}{(1-q)+\nu^{2}(\sin^2 \theta+q\cos^2 \theta)},\\
\gamma  &  =\frac{\nu^{2}\sin^2\theta}{(1-q)+\nu^{2}(\sin^2 \theta+q\cos^2 \theta)},
\end{align*}
respectively, with the parametrized acceleration
$q\equiv e^{-2\pi\Omega/a}$ and the effective coupling
$\nu^{2}\equiv||\lambda||^{2}=\frac{\epsilon^{2}\Omega\Delta}{2\pi}%
e^{-\Omega^{2}\kappa^{2}}$ \cite{Landulfo, wangtian,wald94}, where $\Omega^{-1}\ll\Delta$ is required for the  validity of the above definition. Moreover, the effective coupling should be restricted to $\nu^2\ll1$ for the validity of the perturbative approach applied in this paper. Obviously, $q$ is a monotonous function of the acceleration $a$. In particular, we have $q\rightarrow 0$ for a zero acceleration, and in the limit of infinite acceleration
$q\rightarrow 1$.

\section{Quantifications of  quantum coherence}
Coherence properties of a quantum state are usually attributed to the off-diagonal elements of its density matrix with respect to a selected
reference basis.  With such a fundamental framework, the measure
of coherence for a quantum state $\rho$
in a fixed basis can be defined by measuring the distance between
$\rho$ and its nearest incoherent state. The resource based definition and quantification of quantum coherence was  introduced by Baumgratz \emph{et. al.} \cite{Baumgratz}. They suggested that any information-theoretic measure of quantum coherence $C$ is required to satisfy the following conditions:
\begin{itemize}
\item[(C1)] $C(\rho)=0$ iff $\rho \in \mathcal{I}$.
\item[(C2)] Monotonicity under  incoherent selective measurements on average: $C(\rho)\geq \sum_n p_n C(\rho_n)$, where $\rho_n ={ \hat{\mathcal{K}}_n \rho \hat{{\mathcal{K}}}_n^\dagger } / {p_n}$ and $p_n= tr \left( \hat{\mathcal{K}}_n \rho \hat{{\mathcal{K}}}_n^\dagger \right) $, with $\sum_n \hat{{\mathcal{K}}}_n^\dagger  \hat{\mathcal{K}}_n =I$ and $\hat{\mathcal{K}}_n \mathcal{I} \hat{{\mathcal{K}}}_n^\dagger \subset \mathcal{I}$.
\item[(C3)]Convexity (on-increasing under mixing of states), i.e., $C(\sum_n p_n \rho_n ) \leq \sum_n p_n C(\rho_n)$, for any set of states $\{ \rho_n \}$ and probability distribution $\{ p_n \}$.
\end{itemize}
There are several coherence quantifiers which have been proven to satisfy the above criteria, for instance, the  intuitive \textit{$l_1$} norm,
 relative entropy coherence \cite{Baumgratz}, and trace norm distance  coherence \cite{FidelityTrace}.  The \textit{$l_1$} norm  is defined via the off diagonal elements of the density matrix $\rho$ ~\cite{Baumgratz}
\begin{equation}\label{norm0}
C_{l_1}(\rho)=\sum_{\substack{i,j\\ i\ne j}}|\rho_{i,j}|\;.
\end{equation}
 One can also quantify quantum coherence by the relative entropy
of coherence~\cite{Baumgratz} given by
\begin{equation}\label{entropy0}
C_{RE}(\rho)=S(\rho_{diag})-S(\rho)\;.
\end{equation}
for any state $\rho$, where $S(\rho)=-\textbf{Tr}(\rho\log\rho) $ is the von Neumann entropy and $\rho_{diag}$ is the matrix containing only  diagonal elements of $\rho$ in the reference basis. Alternatively,  the trace norm distance (T-norm)  has been proposed as a coherence measure, which is \begin{equation}
  \label{Eq:Def.Ctr}C_{\text{tr}}(\rho):=\min_{\delta\in \mathcal{I}}\norm{\rho-\delta}_1,
  \end{equation} where $\mathcal{I}$ is a set of incoherent states.

\section{Dynamics behavior of quantum coherence for the detector model}

Our aim is to study the dynamics behaviors of quantum coherence under the Unruh thermal noise  and find under which condition  coherence can be frozen with  the increasing of the
acceleration parameter $q$.  Using Eqs. (\ref{rhof2} -\ref{entropy0}), we can explicitly get the \textit{$l_1$} norm of coherence
\begin{equation}
C_{l_1}(\rho_{t}^{AR})=2 \alpha\sin2\theta \;,
\label{norm1}\end{equation}
 and the relative entropy
of coherence,
\begin{eqnarray}
\nonumber C_{RE}(\rho_{t}^{AR})&=&\sum_{i=1}^3\lambda_i\log_2\Lambda_i
-\beta\log_2\beta-\gamma\log_2\gamma\\ \nonumber&&-2\alpha \sin^2\theta\log_2(2\alpha \sin^2\theta)\\&&-2\alpha \cos^2\theta\log_2(2\alpha cos^2\theta)\;,
\end{eqnarray}
for the detectors' final state density matrix $\rho_{t}^{AR}$, where $\lambda_i (i=1,2,3)$ are nonzero eigenvalues of the final state $\rho_{t}^{AR}$.

To obtain the trace norm  coherence of the final state $\rho_{t}^{AR}$, one needs to seek the nearest incoherence state $\delta^{AR}$ among the set of incoherent states $\mathcal{I}$. Fortunately, we noticed that $\rho_{t}^{AR}$ has non-zero elements only along its diagonal and anti-diagonal. In other words, the state given by Eq. (\ref{rhof2}) describes a
 bipartite system with maximally mixed marginals ($M_{2}^{3}$)~\cite{Horodecki1, Horodecki2}. Therefore, the nearest diagonal incoherence state $\delta^{AR}$  for $\rho_{t}^{AR}$ in
    trace norm is given by $diag(\rho_{t}^{AR})$. The trace norm coherence $C_{\text{tr}}(\rho_{t}^{AR})$ is found to be
\begin{equation}
C_{\text{tr}}(\rho_{t}^{AR})=2 \alpha\sin2\theta,
  \end{equation} which is identical with the \textit{$l_1$} norm of coherence
$C_{l_1}(\rho_{t}^{AR})$.  It have been found in  \cite{FidelityTrace} that the trace norm of coherence for a \emph{one-qubit state } has the same form of expression with the $l_1$ norm. Here we have come to the same conclusion for a two-qubit system in the entangled one-detector-accelerated setting.

\begin{figure*}[tbp]
\centerline{\includegraphics[width= 16cm]{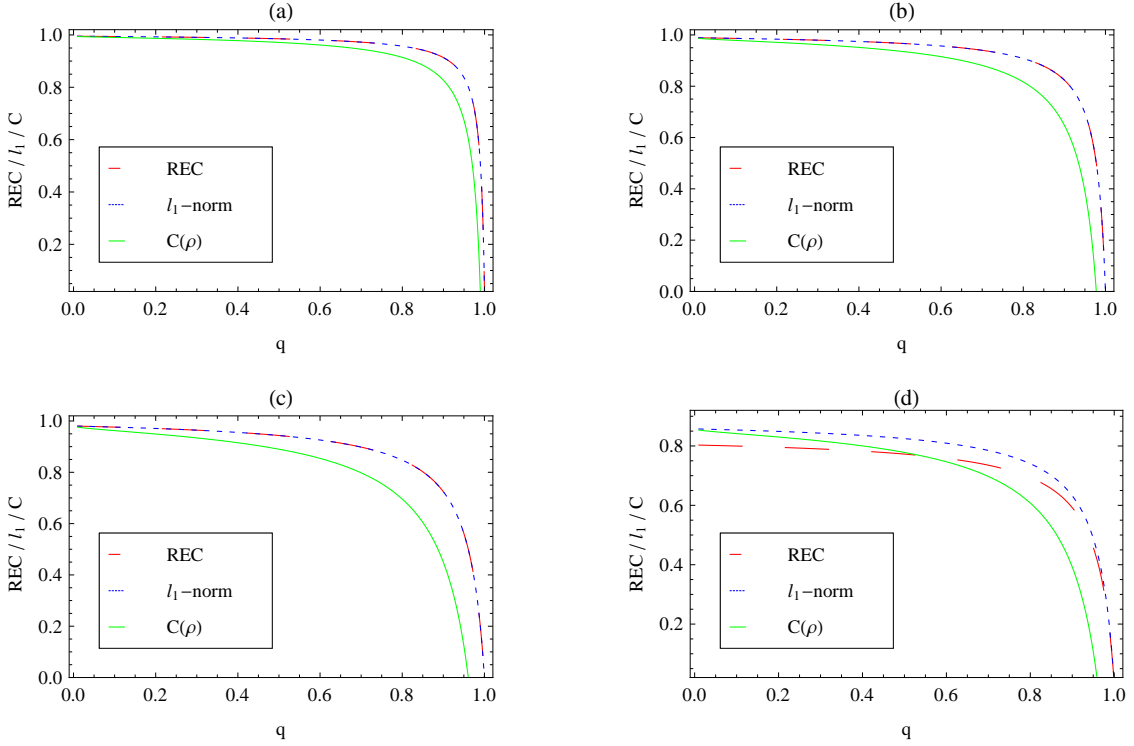}}
\caption{(Color online) Quantum coherence (dashed lines) and entanglement (solid line) between the detectors as a function of the acceleration parameter $q$. The initial state parameter $\theta$ and effective coupling parameter $\nu$ are fixed as (a) $\theta=\pi/4, \nu^2=0.01$, (b) $\theta=\pi/4, \nu^2=0.0225$, (c) $\theta=\pi/4, \nu^2=0.04$, and (d) $\theta=\pi/6, \nu^2=0.04$, respectively.}
\label{Fig1}
\end{figure*}

We are particularly concerned about whether the coherence of the final state can be frozen under some initial or interaction conditions. Such conditions can be obtained by differentiating the coherence of the final state with respect to the acceleration parameter $q$:
$\partial_{q}C(\rho_{t}^{AR})\;$ \cite{thomas,xiaobao}.
That is, coherence is unaffected during the interaction between the accelerated detector and  external scalar field if the differential is zero. By taking the $q$ derivative of the \textit{$l_1$} norm (or the T-norm coherence), we obtain
\begin{equation}
\partial_{q}C_{l_1}(\rho_{t}^{AR})=-\frac{v^2 \sin 2 \theta}{(1- q + q v^2 \cos^2\theta +
   v^2 \sin\theta^2)^2}\;,
\end{equation}
which equals to zero only for $\sin \theta=0$ or $v=0$. The former means that the initial state should be prepared as an incoherence state and the latter indicates that the detectors have no interaction with the external scalar field. We can see that the decrease of quantum coherence can not be frozen under the influence of thermal noise induced by the Unruh effect. The decoherence of the  detectors's quantum
state is  irreversible due to the interaction between the accelerated detector and its external  scalar field.

Now let us discuss this phenomenon further in the theory of open quantum systems \cite{Breuer}.
The evolution from the detectors initial state $|\Psi_{AR}\rangle$ to the final state $\rho_{t}^{AR}$ in Eq. (\ref{rhof2}) can also be represented by
\begin{eqnarray}
\rho_{t}^{AR} =\sum_{\mu \nu} M^{A}_\mu \otimes M^{R}_\nu |\Psi_{AR}\rangle\langle\Psi_{AR}|
(M^{A}_\mu \otimes M^{R}_\nu)^{\dag},
 \end{eqnarray}
where $M_{\mu}^{A}$ and $M_{\mu}^{B}$ are the Kraus operators acting on Alice's and Rob's state. Since Alice's detector keeps static and is switched off, $M_{\mu}^{A}$ is in fact an  identity matrix. After some calculations, we find that Rob's  Kraus operators have the forms
\begin{eqnarray}
\nonumber M_1^{R}&=&\left(\begin{array}{cc}
           \sqrt{1-q}&0\\
           0&\sqrt{1-q}
           \end{array}\right), M_2^{R}=\left(\begin{array}{cc}
           0&0\\
           v\sqrt{q}&0
           \end{array}\right), \\ M^{R}_3 &=&\left(\begin{array}{cc}
                                           0&v\\
                                           0&0
                                          \end{array}\right).
\end{eqnarray}
Applying these Kraus operators to the nearest incoherence state of the initial state $|\Psi_{AR}\rangle$ and normalizing the output state, we obtain
\begin{eqnarray}
 \left(
                  \begin{array}{cccc}
                    \gamma & 0 & 0 & 0 \\
                    0 & 2\alpha \sin^2 \theta  & 0  & 0 \\
                    0 & 0  & 2\alpha\cos^2 \theta  & 0 \\
                    0 & 0 & 0 & \beta \\
                  \end{array}
                \right)
,
\end{eqnarray}
which is exactly the nearest incoherence state of the final state $\rho_{t}^{AR}$. That is to say, the effect of the Unruh radiation acts on the detectors' state as incoherent operations. Therefore, there is no freezing phenomenon of coherence due to the monotonicity of the coherence measures.
\begin{figure*}[tbp]
\centerline{\includegraphics[width= 12cm]{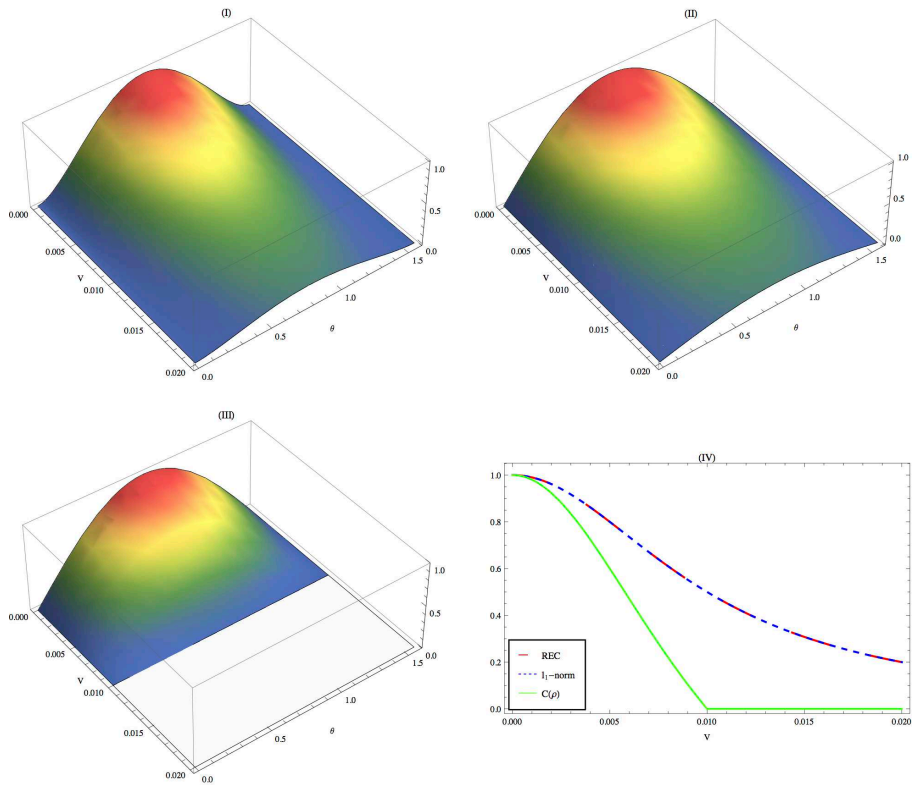}}
\caption{(Color online) Relative entropy coherence (I),  \textit{$l_1$} norm coherence (II), and entanglement (III) between the detectors as  functions of the initial state parameter $\theta$ and the effective coupling parameter $\nu$ for an extremely large acceleration  ($q=0.9999$). (IV) Quantum coherence (dashed lines) and entanglement (solid line) as a function of the effective coupling parameter $\nu$ for an extremely large acceleration. In figure (IV)  the initial state parameter $\theta$ is fixed as $ \theta=\pi/4$.}
\label{Fig2}
\end{figure*}

In Fig. (\ref{Fig1}) we plot the \textit{$l_1$} norm (or the T-norm coherence),  relative entropy
coherence (REC), as well as quantum entanglement of the final state given by Eq. (\ref{rhof2}) as functions of the acceleration parameter $q$ for some fixed coupling parameter $\nu$ and initial state parameter $\theta$. Here we employ the concurrence \cite{Wootters,Coffman} as an  entanglement quantifier, which is given by $C(\rho)=2\max\left\{  0,\tilde{C}_{1}(\rho),\tilde{C}_{2}(\rho)\right\}$
 for a state with a $X$-type structure.
In this definition $\tilde{C}_{1}(\rho)=\sqrt{\rho_{14}\rho_{41}}-\sqrt{\rho_{22}\rho
_{33}}$ and $\tilde{C}_{2}(\rho)=\sqrt{\rho_{23}\rho_{32}}-\sqrt{\rho
_{11}\rho_{44}}$, with $\rho_{ij}$ being  matrix elements of the final state density matrix $\rho_{t}^{AR}$.
The initial state parameter $\theta$ for Figs. (1a, 1b, 1c) is fixed as  $\theta=\pi/4$, i.e., the initial states are singlet states. It is shown that both the quantum coherence and entanglement are monotone degraded with the growth of  the acceleration parameter $q$, which means that the thermal noise induced by Unruh radiation destroys all types of quantum resources. However, the quantum coherence  approaches to zero only in the limit of infinite acceleration
$q\rightarrow 1$, while quantum entanglement could reduce to zero for a \emph{finite} acceleration. In other words, quantum coherence is more robust than entanglement as the Unruh temperature increases.

In Fig. 2 (I-III), we plot the relative entropy coherence (I),  \textit{$l_1$} norm coherence (II), and entanglement (III) between the detectors as  functions of the initial state parameter $\theta$ and the effective coupling parameter $\nu$ for an extremely large acceleration  ($q=0.9999$).  As discussed in
\cite{Schuller-Mann}, this limit describes a physical  picture where Alice is
freely falling into a black hole while the accelerated Rob
barely escapes from the black hole with an extremely large acceleration. We find that both quantum coherence and entanglement monotone decrease with the increases of  effective coupling parameter $\nu$, which means that the interaction between the detector and field destroys quantum resources. It is also shown that, like the behavior of quantum coherence, quantum entanglement of the final state oscillates with the increase of  $\theta$ for any value of $\nu$.
We plot in Fig. (2 IV) the coherence and entanglement as  a function of the coupling parameter $\nu$ for an extremely large acceleration.
We can see that the entanglement decreases more quickly than coherence and suffers from a sudden death as  the effective coupling parameter $\nu$ increases. Comparing with the behavior of quantum coherence versus $q$, quantum entanglement suffers a sudden death for an much smaller coupling parameter. That is to say, quantum coherence is even more robust than entanglement under the influence of the interaction between the accelerated detector and its surrounding field. Then we can safely  arrive at the conclusion that quantum coherence is more robust than entanglement under the effect of Unruh thermal noise and therefore the coherence type quantum resource is more accessible for relativistic quantum information processing tasks.

\section{Conclusions}

In conclusion,  we have studied quantum coherence for two entangled Unruh-Dewitt detectors when one of them is accelerated and interacted with a massless scalar field.  We employ the Unruh-Dewitt detector model, which interacts locally with the neighbor external field. This model avoids  a physically unfeasible detection of global free models in the full space \cite{Landulfo,Landulfo1,tianwang,wangtian}. We find that the quantum coherence can not be frozen during the whole evolution, which is due to the influence of the Unruh thermal noise. It is  shown that  quantum coherence is more robust than entanglement over thermal noise induced by the Unruh effect and therefore the coherence type quantum resources are more accessible for relativistic quantum information processing tasks. We known that an accelerated observer in the Minkowski vacuum corresponds to a static
observers outside a black hole
in the Hartle-Hawking vacuum \cite{Unruh,Schuller-Mann,Landulfo1}. Similarly, a static observer in the Minkowski space-time corresponds to  a free-falling observer  in the Schwarzschild spacetime. Therefore, the analysis used to derive the results of our manuscript can, in principle, be applied to study  the dynamic behavior of quantum coherence under the influence of Hawking radiation.

\begin{acknowledgments}
J. W. is supported the National Natural Science Foundation
of China under Grant No. 11305058, the Doctoral Scientific Fund Project of the Ministry of Education of China under Grant No. 20134306120003, and the Postdoctoral Science Foundation of China under Grant No. 2014M560129 and No. 2015T80146. J. J. is supported the National Natural Science Foundation
of China under Grant No. 11475061. H. F.  is supported the National Natural Science Foundation
of China under Grant No. 91536108.

\end{acknowledgments}

\end{document}